\documentclass[aps,pra,twocolumn,groupedaddress]{revtex4-1}
\usepackage{bm,amsmath,graphicx,color}	
\begin{document}
\title{Enhanced second harmonic generation and photon drag effect in a doped graphene placed on a two-dimensional diffraction grating}
\author{Tetsuyuki Ochiai}
\affiliation{Research Center for Functional Materials, National Institute for Materials Science (NIMS), Tsukuba 305-0044, Japan}
\begin{abstract}
We theoretically investigate the second harmonic generation and photon drag effect induced by an incident plane wave to a doped graphene placed on a two-dimensional diffraction grating. The relevant nonlinear conductivity of the graphene is obtained by a semi-classical treatment with a phenomenological relaxation. The grating acts not only as a plasmon coupler but also as a dispersion modulator of the graphene plasmon.  As a result, the second harmonic generation is strongly enhanced by exciting the graphene plasmon polariton of the first- and/or second-harmonic frequencies. The photon drag effect is also strongly enhanced by the excitation of the plasmon at the first-harmonic frequency.  The direct current induced by the photon drag effect flows both forward and backward directions to the incident light, depending on the modulated plasmon mode concerned.  
 
\end{abstract}
\date{\today}
\maketitle

\section{Introduction}

Graphene is a monolayer of Carbon atoms arranged in a honeycomb lattice, and exhibits various interesting physical properties \cite{RevModPhys.81.109}. Its electronic band structure consists of conically touching valence and conduction bands (called the Dirac cone) at the corners of the first Brillouin zone. This band structure with the Fermi energy lying at the contact point (the Dirac point), provides unique features in various aspects, such as quantum transport \cite{novoselov2005tdg,zhang2005experimental}, nonlinear optics \cite{PhysRevB.82.201402,hendry2010coherent}, magnetism \cite{yazyev2010emergence}, spintronics \cite{han2014graphene}, etc.

Although this ``bare'' graphene is an interesting object in physics, chemistry, and engineering, a further application of graphene is available by carrier doping or electric gating. 
If it is carrier doped or electrically gated, the Fermi energy shifts from the Dirac point, and there are free carriers at the Fermi energy, like as in metal. In such a system, the plasmon, collective excitation of carriers, plays an important role. 

In the doped graphene, the plasmon is typically in the THz frequency range depending on the amount of the doping, and has the $\sqrt{k}$ dispersion ($k$ being the momentum parallel to the graphene sheet) in the non-retarded and local-response approximations \cite{bludov2013primer}. Since the $\sqrt{k}$ dispersion curve in the frequency-momentum space is far away at large  $k$ from the light line with the $k$-linear dispersion, the radiation field of the graphene plasmon is tightly confined in the graphene sheet. This confinement results in a strong light-matter interaction in the vicinity of the graphene, giving rise to various applications in the context of graphene plasmonics \cite{koppens2011graphene}.

Without some spatial processing of the graphene, the plasmon cannot couple to the far-field radiation, because the dispersion relation of the plasmon is outside the light cone.  Many optical applications arise when the plasmon couples to the far-field radiation inside the light cone. The attenuated total reflection \cite{fahrenfort1961attenuated} and grating coupling \cite{dakss1970grating} are two typical setups for this purpose.

Here, we focus on the latter coupling, because the grating acts not only as a plasmon coupler to the far-field radiation but also as a dispersion modulator of the plasmon.  In fact, the author showed that the graphene plasmon exhibits a plasmonic band-gap structure if the doped graphene is placed in the vicinity of a two-dimensional (2D) diffraction grating \cite{ochiai2015spatially}. The modulated plasmon dispersion exhibits slow-light effects and enhanced optical density of states. Since the optical density of states is the heart of various light-matter interactions via Fermi's golden rule, the modulated plasmon dispersion plays versatile roles. Nonlinear optics is not an exception of the light-matter interactions.

In this paper, we study the enhancement of the second-harmonic generation (SHG), namely the generation of the frequency-doubled light, and photon drag effect (PDE), namely the direct current generation by irradiating light,  as consequences of the second-order optical nonlinearity  in the doped graphene placed  on a two-dimensional diffraction grating. Using a semi-classical approach with a corresponding principle, the second-order optical conductivity is obtained in a concise form. 
The nonlinear conductance together with the graphene plasmon modulated by the grating  induces both the enhanced SHG and PDE in the doped graphene on the grating.  
We evaluate the enhancement factors of these effects relative to those  in the free-standing graphene. We obtain a strong enhancement of the SHG and PDE if the frequency and momentum match to the plasmon polariton.

A similar semi-classical approach is employed to investigate the SHG in modulated graphene (graphene metasurfaces) \cite{manzoni2015second,PhysRevB.92.161406}. As for unmodulated graphene,  intensive investigation on the SHG and PDE has been carried out \cite{PhysRevB.81.165441,ivchenko2012photoinduced,PhysRevB.90.241416,glazov2014high,PhysRevB.92.235307}. The former neglects dissipation, which is needed to formulate the PDE even in the clean limit. Various approaches employed in the latter have difficulty in applying to modulated graphene structures. Therefore, our approach provides a convenient way to analyze both the SHG and PDE in a modulated graphene, in an equal footing.   
Besides, in metallic nanostructures such as metal-coated gratings, surface- or particle-plasmon enhanced SHG or PDE have been investigated \cite{PhysRevB.24.849,nahata2003enhanced,airola2005second,PhysRevLett.103.103906,kurosawa2012surface,noginova2013plasmon,PhysRevLett.117.083901}. Our work can be also regarded as an extension of these works to the graphene plasmon.

This paper is organized as follows. In Sec. II, we derive the second-order optical nonlinearity in the doped graphene.  We then formulate the SHG and PDE in the doped graphene placed on a diffraction grating in Sec. III.  
Numerical results and discussions are given in Sec. IV. Finally, the conclusion is given in Sec. V.

\section{Nonlinear conductivity of doped graphene}

Let us consider the nonlinear optical conductivity of the doped (or gated) graphene via a semi-classical approach. We employ a classical non-relativistic treatment of charged particles \cite{PhysRevLett.117.083901} and translate the results by a ``corresponding principle'' for the Dirac spectrum of the graphene.

Suppose that a carrier of mass $m$ and charge $e$ moves in the 2D plane under the external radiation field. Its dynamics is governed by 
\begin{align}
m\left(\frac{{\rm d}^2{\bm r}_\alpha}{{\rm d}t^2} +\frac{1}{\tau}\frac{{\rm d}{\bm r}_\alpha}{{\rm d}t}\right) = e \left({\bm E}({\bm r}_\alpha,t) + \frac{{\rm d}{\bm r}_\alpha}{{\rm d}t}\times {\bm B}({\bm r}_\alpha,t)\right),
\end{align}
with a phenomenological relaxation time $\tau$. 
We solve the equation perturbatively with respect to the external field. 
We thus put the carrier position ${\bm r}_\alpha={\bm r}_\alpha^{(0)} +{\bm r}_\alpha^{(1)} + {\bm r}_\alpha^{(2)} +\dots$, being 
${\bm r}_\alpha^{(i)}$ the $i$-th order term of the external field. 

In the zero-th order, the equation becomes simply as  
\begin{align}
m\left(\frac{{\rm d}^2{\bm r}_\alpha^{(0)}}{{\rm d}t^2} +\frac{1}{\tau} \frac{{\rm d}{\bm r}_\alpha^{(0)}}{{\rm d}t}\right) =0,
\end{align}
resulting in 
\begin{align}
{\bm r}_\alpha^{(0)}={\bm r}_0 + {\bm v}_0 \tau (1-{\rm e}^{-\frac{t}{\tau}}). 
\end{align}
Since the initial velocity is diversely distributed,  
we put ${\bm r}_\alpha^{(0)}\to {\bm r}_0$. 

In the first order, the equation becomes 
\begin{align}
m\left(\frac{{\rm d}^2{\bm r}_\alpha^{(1)}}{{\rm d}t^2} +\frac{1}{\tau} \frac{{\rm d}{\bm r}_\alpha^{(1)}}{{\rm d}t}\right) = e {\bm E}({\bm r}_0,t).  
\end{align} 
Suppose that the external field is monochromatic with (angular) frequency $\omega$, then we have  
\begin{align}
&{\bm E}({\bm r},t)=\Re[{\bm E}_\omega ({\bm r}){\rm e}^{-{\rm i}\omega t}],\\
&{\bm r}_\alpha^{(1)}=\Re[{\bm r}_\omega^{(1)}{\rm e}^{-{\rm i}\omega t}],\\
&  {\bm r}_\omega^{(1)}= -\frac{e}{m\omega(\omega + {\rm i}\tau^{-1})} {\bm E}_\omega ({\bm r}_0).
\end{align}
Here, $\Re$ stands for the real part.

In the second order, the equation becomes 
\begin{align}
&m\left(\frac{{\rm d}^2{\bm r}_\alpha^{(2)}}{{\rm d}t^2} +\frac{1}{\tau} \frac{{\rm d}{\bm r}_\alpha^{(2)}}{{\rm d}t}\right) \nonumber \\
&\hskip20pt = e \left({\bm r}_\alpha^{(1)}\cdot{\bm \nabla}{\bm E}({\bm r},t)|_{{\bm r}_0} + \frac{{\rm d}{\bm r}_\alpha^{(1)}}{{\rm d}t}\times {\bm B}({\bm r}_0,t)\right).
\end{align}
We can show that the velocity has the DC and AC components as  
\begin{align}
&\frac{{\rm d}{\bm r}_\alpha^{(2)}}{{\rm d}t}={\bm v}_{\rm DC}^{(2)} + {\bm v}_{\rm AC}^{(2)}(t),\\
&{\bm v}_{\rm DC}^{(2)}=\frac{e\tau}{2m}\Re\left[{\bm r}_\omega^{(1)}{}^* \cdot {\bm \nabla}{\bm E}_\omega ({{\bm r}_0}) + {\bm r}_\omega^{(1)}{}^* \times {\bm \nabla}\times {\bm E}_\omega ({{\bm r}_0}) \right],\\
&{\bm v}_{\rm AC}^{(2)}(t)=\Re[{\bm v}_{2\omega}^{(2)}{\rm e}^{-2{\rm i}\omega t}],\\
&{\bm v}_{2\omega}^{(2)}=\frac{{\rm i}e}{2m(2\omega + {\rm i}\tau^{-1})} \nonumber \\
&\hskip20pt \times \left(
{\bm r}_\omega^{(1)} \cdot {\bm \nabla}{\bm E}_\omega ({{\bm r}_0}) - {\bm r}_\omega^{(1)} \times {\bm \nabla}\times {\bm E}_\omega ({{\bm r}_0}) 
\right).
\end{align}

The charge and current densities  of multi-particle systems (each particle labeled by $\alpha$) are expressed as 
\begin{align}
&\rho({\bm r},t)=e\sum_\alpha \delta({\bm r}-{\bm r}_\alpha(t)),\\
&{\bm j}({\bm r},t)= e\sum_\alpha \frac{{\rm d}{\bm r}_\alpha(t)}{{\rm d}t}\delta({\bm r}-{\bm r}_\alpha(t)), 
\end{align} 
where ${\bm r}$ is the coordinate in 2D plane, and $\delta({\bm r})$ is the Dirac's delta function. 
We follow the same procedure of the perturbation with respect to the external field: $\rho=\rho^{(0)} + \rho^{(1)} + \rho^{(2)} + \dots$ 
and ${\bm j}={\bm j}^{(0)}+{\bm j}^{(1)}+{\bm j}^{(2)}+\dots$.  
In the zero-th order, we assume that the macroscopic distribution is uniform, namely, $\rho^{(0)}=\rho_0(=en)$ and ${\bm j}^{(0)}=0$. 
In the first order, we have the well-known Drude form of the local response as  
\begin{align}
&{\bm j}^{(1)}({\bm r},t)=\Re[{\bm j}_\omega^{(1)}({\bm r}){\rm e}^{-{\rm i}\omega t}],\\
&{\bm j}_\omega^{(1)}({\bm r}) = \sigma_\omega^{(1)} {\bm E}_\omega({\bm r}),\\
& \sigma_\omega^{(1)}= \frac{{\rm i}e\rho_0}{m(\omega + {\rm i}\tau^{-1})}. \label{Eq_drude0}
\end{align}
In the second order, the current density becomes 
\begin{align}
&{\bm j}^{(2)}({\bm r},t)={\bm j}_{\rm DC}^{(2)}({\bm r})+{\bm j}_{\rm AC}^{(2)}({\bm r},t),\\
& [{\bm j}_{\rm DC}^{(2)}({\bm r})]_i=-\frac{e^2\rho_0\tau}{2m^2\omega}\Re\left[\frac{1}{\omega-{\rm i}\tau^{-1}} \left({\bm E}_\omega^* \cdot {\bm \nabla}{\bm E}_\omega  + {\bm E}_\omega^* \times {\bm \nabla}\times {\bm E}_\omega  \right)_i\right] \nonumber \\
&\hskip45pt + \frac{e^2\rho_0}{2m^2\omega(\omega^2+\tau^{-2})}\partial_k\Im[({\bm E}_\omega^*)_i ({\bm E}_\omega)_k],\\
&{\bm j}_{\rm AC}^{(2)}({\bm r},t)=\Re[{\bm j}_{2\omega}^{(2)}({\bm r}){\rm e}^{-2{\rm i}\omega t}],\\
&[{\bm j}_{2\omega}^{(2)}({\bm r})]_i=-\frac{{\rm i}e^2\rho_0}{2m^2\omega(\omega+{\rm i}\tau^{-1} )(2\omega + {\rm i}\tau^{-1} )} \nonumber \\
&\hskip70pt \times \left(  
{\bm E}_\omega  \cdot {\bm \nabla}{\bm E}_\omega  -  {\bm E}_\omega  \times {\bm \nabla}\times {\bm E}_\omega  \right)_i \nonumber \\
&\hskip45pt +\frac{{\rm i}e^2\rho_0}{2m^2\omega (\omega+{\rm i}\tau^{-1} )^2} \partial_k [({\bm E}_\omega)_i({\bm E}_\omega)_k],   
\end{align}
where $\Im$ stands for the imaginary part, and repeated index of $k$ is summed 
up [$k=1(x)$ and $2(y)$]  in accordance with Einstein's convention. 
The DC components gives rise to the PDE, while the AC components yields the SHG.

Besides, the linear conductivity of the doped graphene obtained from the Dirac spectrum  is generally written as \cite{bludov2013primer}  
\begin{align}
\sigma_\omega^{(1)} =\frac{{\rm i}e^2E_{\rm F}}{\pi\hbar^2}\frac{1}{\omega + {\rm i}\tau^{-1}},  \label{Eq_drude}
\end{align}
where $E_{\rm F}$ is the Fermi energy measured from the Dirac point, and the interband transition is neglected assuming a classical regime, $\hbar\omega \ll E_{\rm F}$.
This expression is obtained from Eq. (\ref{Eq_drude0}) with the following replacement (``corresponding principle'') of carrier density $n$ and mass $m$ in accordance with the Dirac spectrum:  
\begin{align}
&n \to g_{\rm s}g_{\rm v}\int \frac{{\rm d}^2k}{(2\pi)^2}\theta(k_{\rm F}-k)=\frac{k_{\rm F}^2}{\pi},\\
&m \to \frac{\hbar k_{\rm F}}{v_{\rm F}}. 
\end{align}
Here, $g_{\rm s}(=2)$ and $g_{\rm v}(=2)$ are the degree of spin and valley degeneracies, respectively, and  $E_{\rm F}=\hbar v_{\rm F}k_{\rm F}$ being $v_{\rm F}(\simeq c/300)$ the effective velocity of the Dirac spectrum.

If we apply this principle to the second-order optical nonlinearity, we obtain 
\begin{align}
&[{\bm j}_{\rm DC}^{(2)}({\bm r})]_i=-\frac{e^3 v_{\rm F}^2}{2\pi \hbar^2}\frac{\tau}{\omega}\Re\left[\frac{1}{\omega-{\rm i}\tau^{-1}} ({\bm E}_\omega)_k^*\partial_i ({\bm E}_\omega)_k\right] \nonumber \\
&\hskip45pt + \frac{e^3 v_{\rm F}^2}{2\pi\hbar^2}\frac{1}{\omega(\omega^2+\tau^{-2})}
\partial_k\Im[({\bm E}_\omega)_i^*({\bm E}_\omega)_k],\label{Eq_PDE_current}\\
&[{\bm j}_{2\omega}^{(2)}({\bm r})]_i=\frac{e^3 v_{\rm F}^2}{2\pi {\rm i}\hbar^2}\frac{1}{\omega(\omega + {\rm i}\tau^{-1})(2\omega+{\rm i}\tau^{-1})} \nonumber \\
&\hskip70pt \times \left[2({\bm E}_\omega)_k\partial_k ({\bm E}_\omega)_i - ({\bm E}_\omega)_k\partial_i ({\bm E}_\omega)_k\right]  \nonumber \\
&\hskip40pt - \frac{e^3 v_{\rm F}^2}{2\pi {\rm i}\hbar^2}\frac{1}{\omega(\omega+{\rm i}\tau^{-1})^2}
\partial_k\left[({\bm E}_\omega)_i({\bm E}_\omega)_k\right].\label{Eq_SHG_current}
\end{align}
The result is independent of the doping $E_{\rm F}$, and characterized by the dimensional factor of $e^3v_{\rm F}^2/\hbar^2$.

In the semi-classical approach via the Boltzmann equation given in Ref. \onlinecite{manzoni2015second}, the Lorentz-force term and dissipation are neglected, resulting in a difference in the AC nonlinear conductivity from Eq. (\ref{Eq_SHG_current}). However, the dimensional factor $e^3 v_{\rm F}^2/\hbar^2$ is common between the two approaches.  If we introduce the phenomenological relaxation time $\tau$ to the Boltzmann equation in a naive way, the current conservation is no longer fulfilled \cite{kragler1980dielectric}. Since the dissipation becomes important in the PDE, the approach fails to explain possible PDE in the doped graphene under a spatial modulation.  Although, our approach does not include faithfully 
the Dirac spectrum and quantum nature of the doped graphene,  we suppose that 
physics essential in the SHG and PDE of the doped graphene can be grasped with this approach.

More accurate approaches such as in Refs. \onlinecite{glazov2014high,PhysRevB.92.235307}  yield complex (in formula) and numerical conductivity. Therefore, the merging with Maxwell-equation solvers such as the finite-difference time-domain method and rigorous coupled-wave analysis seems to be difficult.

\section{Light scattering by doped graphene on grating}

\begin{figure}
\centerline{
\includegraphics[width=0.4\textwidth]{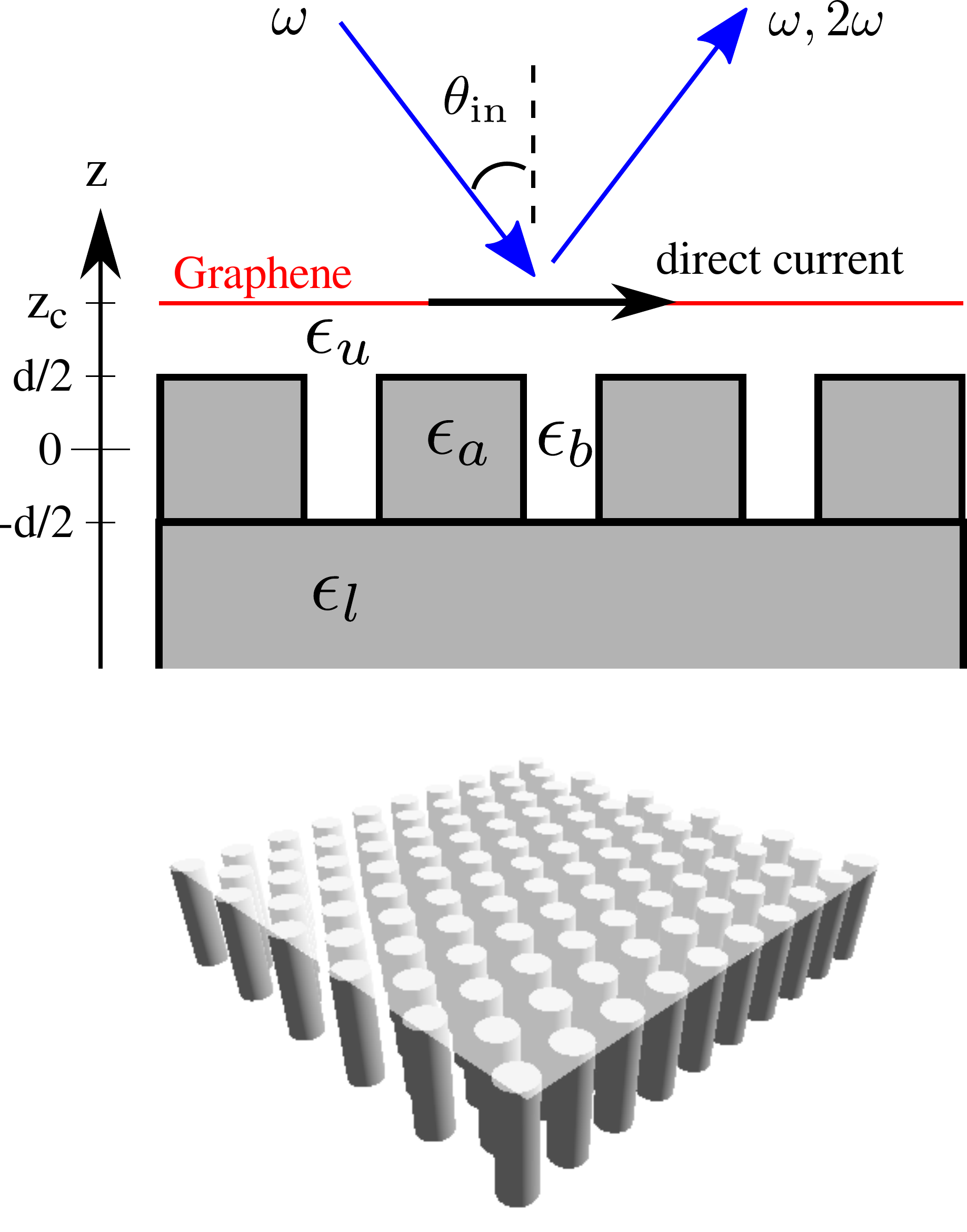}}
\caption{\label{Fig_setup}Schematic illustration of the system under study. Doped graphene is placed on a triangular-lattice diffraction grating composed of circular cylinders.  The incident light is coming from the top of the structure, and  the second-harmonic light and direct current are generated.  }
\end{figure}

Next, let us consider the light scattering problem of the graphene-grating system. 
The system under study is shown in Fig. \ref{Fig_setup}. The doped graphene is placed on a two-dimensional diffraction grating, and a plane-wave light is incident from the top of the structure. 
The SHG light and the PDE current are induced.

As predicted in Ref. \onlinecite{ochiai2015spatially}, the linear optical conductivity of the doped graphene placed on a diffraction grating exhibits the spatially periodic modulation depending on the grating geometry, distance between the graphene and grating, and the Fermi energy. 
Taking account of the grating periodicity, the modulated linear conductivity is expressed as 
\begin{align}
\sigma({\bm r};\omega)=\sum_{\bm g}e^{i{\bm g}\cdot{\bm r}}\sigma_{\bm g}(\omega),
\end{align} 
where ${\bm g}$ is a reciprocal lattice vector of the grating.

In the perturbation viewpoint, the second-harmonic (SH) wave is generated through the second-order nonlinear current induced in the graphene layer. Suppose that a plane-wave light of in-plane wave number ${\bm k}$ and (angular) frequency $\omega$ is incident on the graphene. This first-harmonic (FH) wave is scattered by the graphene-grating system, and the nonzero in-plane component of the electric field is induced in the graphene. The nonlinear current of Eq. (\ref{Eq_SHG_current}) is then induced, and becomes the source of the SH wave.
The radiation field of the SH wave is obtained by solving the Maxwell equation  with the following boundary condition at the graphene layer:
\begin{align}
&\Delta ({\bm E}_{2\omega})_\|=0,\\ 
&\Delta ({\bm H}_{2\omega})_\|=\left[({\bm j}_{2\omega})_y,-({\bm j}_{2\omega})_x\right],\\
&{\bm j}_{2\omega}({\bm r})={\bm j}_{2\omega}^{(1)}({\bm r}) + {\bm j}_{2\omega}^{(2)}({\bm r}),\label{Eq_2w_current}
\end{align}
where the radiation field of the SH wave is expressed as 
\begin{align}
&{\bm F}({\bm x},t)=\Re[{\bm F}_{2\omega}({\bm x}){\rm e}^{-2{\rm i}\omega t}] \quad ({\bm F}={\bm E},{\bm H}),
\end{align}
and $\Delta$ is the difference between the fields above and below the graphene layer:
\begin{align}
&\Delta F = F({\bm r},z_c+\delta) - F({\bm r},z_c-\delta).   
\end{align} 
with infinitesimal $\delta$. 
Now, the first term in Eq. (\ref{Eq_2w_current}) is the linear current given by 
\begin{align}
{\bm j}_{2\omega}^{(1)}({\bm r})=\sigma({\bm r};2\omega){\bm E}_{2\omega}({\bm r},z_c),
\end{align}
where ${\bm E}_{2\omega}({\bm x})$ is to be determined. 
The second term in Eq. (\ref{Eq_2w_current}) is the  nonlinear current given by Eq. (\ref{Eq_SHG_current}) with ${\bm E}_\omega ({\bm x})$ being the solution of the FH wave, induced in the graphene sheet.  The solution can be obtained with the S-matrix formalism \cite{ochiai2015spatially}. The nonlinear current is expressed as 
\begin{align}
{\bm j}_{2\omega}^{(2)}({\bm r})=\sum_{\bm g}{\bm j}_{\bm g}^{(2)}{\rm e}^{{\rm i}(2{\bm k}+{\bm g})\cdot {\bm r}},  
\end{align}
by the translational invariance.

The SH wave can be expanded as (see Fig. \ref{Fig_setup})
\begin{align}
&{\bm E}_{2\omega}({\bm x})=\sum_{\bm g}{\rm e}^{{\rm i}(2{\bm k}+{\bm g})\cdot{\bm r}}{\bm E}_{2\omega}(z;{\bm g}),\\
&{\bm E}_{2\omega}(z;{\bm g})=\left\{
\begin{array}{ll}
(t_{{\rm p}{\bm g}}^+ {\bm p}_{\bm g}^{{\rm u}+} + t_{{\rm s}{\bm g}}^+{\bm s}_{\bm g})
{\rm e}^{{\rm i}\gamma_{\bm g}^{\rm u}(z-z_c)}       & z>z_c\\
(a_{{\rm p}{\bm g}}^+ {\bm p}_{\bm g}^{{\rm u}+} + a_{{\rm s}{\bm g}}^+{\bm s}_{\bm g})
{\rm e}^{{\rm i}\gamma_{\bm g}^{\rm u}(z-z_c)}   & \\
+(a_{{\rm p}{\bm g}}^- {\bm p}_{\bm g}^{{\rm u}-} + a_{{\rm s}{\bm g}}^- {\bm s}_{\bm g})
{\rm e}^{-{\rm i}\gamma_{\bm g}^{\rm u}(z-z_c)}   & \frac{d}{2} <z<z_c \\
(t_{{\rm p}{\bm g}}^- {\bm p}_{\bm g}^{{\rm l}-} + t_{{\rm s}{\bm g}}^-{\bm s}_{\bm g})
{\rm e}^{-{\rm i}\gamma_{\bm g}^{\rm l}(z+\frac{d}{2})}       & z<-\frac{d}{2}
\end{array}\right.,\label{Eq_SHG_E}
\end{align}
where 
\begin{align}
&{\bm p}_{\bm g}^{u\pm}=\pm \frac{{\gamma}_{\bm g}^{u}}{q_{u}}\widehat{(2{\bm k}_{\bm g})}_\| - \frac{|2{\bm k}_{\bm g}|}{q_{u}}\hat{z},\quad 
{\bm s}_{\bm g}= \widehat{(2{\bm k}_{\bm g})}_\perp,\\ 
& 2{\bm k}_{\bm g}=2{\bm k}+{\bm g},\quad q_u=\frac{2\omega}{c}\sqrt{\epsilon_u},\\
&\gamma_{\bm g}^u=\sqrt{q_u^2-|2{\bm k}_{\bm g}|^2},\\
&\hat{\bm k}_\|=\frac{1}{|{\bm k}|}(k_x,k_y), \quad \hat{\bm k}_\perp=\frac{1}{|{\bm k}|}(-k_y,k_x). 
\end{align}
Using the S-matrix of the diffraction grating at $2\omega$, the unknown coefficients $t_{{\rm p}{\bm g}}^\pm$,  $t_{{\rm s}{\bm g}}^\pm$, $a_{{\rm p}{\bm g}}^\pm$, and  $a_{{\rm s}{\bm g}}^\pm$
can be solved as
\begin{align}
& \sum_{{\bm g}'}\left\{
\hat{1}\delta_{{\bm g}{\bm g}'}-[\tilde{S}^{+-}(-\sigma_3+\tilde{S}^{+-})^{-1}]_{{\bm g}{\bm g}'}
-[\sigma_3(-\sigma_3+\tilde{S}^{+-})^{-1}]_{{\bm g}{\bm g}'} \right.\nonumber \\
& \hskip10pt \left. +\frac{c\mu_0}{\sqrt{\epsilon_u}}
\sigma_{{\bm g}-{\bm g}'}^{(1)}\left[
\begin{array}{cc}
 \frac{\gamma_{{\bm g}'}^u}{q_u}
(\widehat{2{\bm k}_{\bm g}})_\perp\cdot
(\widehat{2{\bm k}_{{\bm g}'}})_\perp 
& -(\widehat{2{\bm k}_{\bm g}})_\perp\cdot
(\widehat{2{\bm k}_{{\bm g}'}})_\| \\
\frac{\gamma_{{\bm g}'}^u}{\gamma_{\bm g}^u}
(\widehat{2{\bm k}_{\bm g}})_\|\cdot
(\widehat{2{\bm k}_{{\bm g}'}})_\perp 
& -\frac{q_u}{\gamma_{\bm g}^u}
(\widehat{2{\bm k}_{\bm g}})_\|\cdot
(\widehat{2{\bm k}_{{\bm g}'}})_\|
\end{array}\right]\right\} \nonumber \\
&\hskip40pt 
\times \left(\begin{array}{c}
t_{{\rm p}{\bm g}'}^+\\
t_{{\rm s}{\bm g}'}^+
\end{array}\right) =\frac{c\mu_0}{\sqrt{\epsilon_u}}\left(
\begin{array}{c}
-(\widehat{2{\bm k}_{\bm g}})_\|\cdot {\bm j}_{\bm g}^{(2)}\\
\frac{q_u}{\gamma_{\bm g}^u}
(\widehat{2{\bm k}_{\bm g}})_\perp\cdot {\bm j}_{\bm g}^{(2)}
\end{array}\right),\\
& \left(\begin{array}{c}
a_{{\rm p}{\bm g}}^+\\
a_{{\rm s}{\bm g}}^+
\end{array}\right) = \sum_{{\bm g}'}
[\tilde{S}^{+-}(-\sigma_3+\tilde{S}^{+-})^{-1}]_{{\bm g}{\bm g}'}
\left(\begin{array}{c}
t_{{\rm p}{\bm g}'}^+\\
t_{{\rm s}{\bm g}'}^+
\end{array}\right),\\
& \left(\begin{array}{c}
a_{{\rm p}{\bm g}}^-\\
a_{{\rm s}{\bm g}}^-
\end{array}\right) = \sum_{{\bm g}'}
[(-\sigma_3+\tilde{S}^{+-})^{-1}]_{{\bm g}{\bm g}'}
\left(\begin{array}{c}
t_{{\rm p}{\bm g}'}^+\\
t_{{\rm s}{\bm g}'}^+
\end{array}\right),\\
& \left(\begin{array}{c}
t_{{\rm p}{\bm g}}^-\\
t_{{\rm s}{\bm g}}^-
\end{array}\right) = \sum_{{\bm g}'}
[\tilde{S}^{--}]_{{\bm g}{\bm g}'}
\left(\begin{array}{c}
a_{{\rm p}{\bm g}'}^-\\
a_{{\rm s}{\bm g}'}^-
\end{array}\right). 
\end{align}
Here, the S-matrix $\tilde{S}$ is defined by 
\begin{align}
\left(\begin{array}{c}
a_{{\rm p}{\bm g}}^+ \\
a_{{\rm s}{\bm g}}^+ \\
t_{{\rm p}{\bm g}}^- \\
t_{{\rm s}{\bm g}}^- 
\end{array}\right)= \sum_{{\bm g}'}\left(\begin{array}{cc}
(\tilde{S}^{++})_{{\bm g}{\bm g}'} & (\tilde{S}^{+-})_{{\bm g}{\bm g}'} \\
(\tilde{S}^{-+})_{{\bm g}{\bm g}'} & (\tilde{S}^{--})_{{\bm g}{\bm g}'}
\end{array}\right)  \left(\begin{array}{c}
d_{{\rm p}{\bm g}'}^+ \\
d_{{\rm s}{\bm g}'}^+ \\
a_{{\rm p}{\bm g}'}^- \\
a_{{\rm s}{\bm g}'}^- 
\end{array}\right),
\end{align}
being $d_{{\rm p(s)}{\bm g}}^+$ the coefficient of the incoming wave of the P(S) polarization from the bottom (substrate), which is absent in the current setup.    
Each $\tilde{S}^{\pm\pm}$ is a $2N_{\bm g}\times 2N_{\bm g}$ matrix with $N_{\bm g}$ being the number of reciprocal lattice vectors taken into account in the numerical calculation.   
The S-matrix is available with the rigorous coupled-wave analysis \cite{Noponen1994,Li1997,Tikhodeev:Y:M:G:I::66:p045102:2002}. 
The intensity of the SHG light in the specular reflected direction is given by 
$\sqrt{|t_{{\rm p}{\bm g}=0}^+|^2 + |t_{{\rm s}{\bm g}=0}^+|^2}$.

Besides, the direct current of the PDE given by Eq. (\ref{Eq_PDE_current}) is expressed as 
\begin{align}
{\bm j}_{\rm DC}^{(2)}({\bm r})=\sum_{\bm g}{\bm j}_{{\rm DC},{\bm g}}^{(2)}{\rm e}^{{\rm i}{\bm g}\cdot{\bm r}}.  
\end{align}
We are interested in the average current density of the PDE per unit cell, which is equal to  ${\bm j}_{{\rm DC},{\bm g}=0}^{(2)}$.

For comparison, we summarize the SHG and PDE in a free-standing doped graphene. 
Suppose that a plane-wave light is incident on the graphene from the top. 
Its electric field is given by 
\begin{align}
&{\bm E}_{\omega}^0({\bm x})={\rm e}^{{\rm i}({\bm k}\cdot{\bm r}-\gamma z)} (d_{\rm p}{\bm p}^- + d_{\rm s}{\bm s}),\\
&{\bm p}^\pm =\pm \frac{\gamma}{q}\hat{\bm k}_\| - \frac{|{\bm k}|}{q}\hat{z},\quad {\bm s}=\hat{\bm k}_\perp, \\
&q=\frac{\omega}{c}, \quad \gamma=\sqrt{q^2-{\bm k}^2}. 
\end{align}
The electric field in the transmitted side is expressed as  
\begin{align}
&{\bm E}_\omega({\bm x})={\rm e}^{{\rm i}({\bm k}\cdot{\bm r}-\gamma z)} (t_{\rm p}{\bm p}^- + t_{\rm s}{\bm s}),\\
&t_{\rm p}=\frac{1}{1+\frac{\gamma}{q}\alpha_e \tilde{\sigma}_{\omega}^{(1)}} d_{\rm p},\\
&t_{\rm s}=\frac{1}{1+\frac{q}{\gamma}\alpha_e \tilde{\sigma}_{\omega}^{(1)}} d_{\rm s},
\end{align}
where  $\alpha_e$ is the fine structure constant and $\tilde{\sigma}_{\omega}^{(1)}$ is the normalized conductance defined by $\sigma_{\omega}^{(1)}=(e^2/h) \tilde{\sigma}_{\omega}^{(1)}$. 
The optical absorption $A$ is measured by 
\begin{align}
A = 2\frac{q}{\gamma} \alpha_e \Re[\tilde{\sigma}_\omega^{(1)}] \frac{ 
|t_{\rm p}|^2 \frac{\gamma^2}{q^2} + |t_{\rm s}|^2}{|d_{\rm p}|^2+|d_{\rm s}|^2 }, 
\end{align}
where $A=0$ in the clean limit ($\tau=\infty$) and $A=1$ for the perfect absorption.

The source current of the SHG and the direct current of the PDE in the graphene are given by 
\begin{align}
&{\bm j}_{2\omega}^{(2)}({\bm r})={\bm j}_{2\omega,2{\bm k}}^{(2)}{\rm e}^{2{\rm i}{\bm k}\cdot{\bm r}},\\
&{\bm j}_{2\omega,2{\bm k}}^{(2)}=
\frac{e^3v_{\rm F}^2}{2\pi\hbar^2}\frac{1}{\omega(\omega+{\rm i}\tau^{-1})} \nonumber\\
&\hskip5pt \times \left( \frac{1}{2\omega+{\rm i}\tau^{-1}} (2{\bm e}_\| ( {\bm k}\cdot{\bm e}_\|) -{\bm k}{\bm e}_\|^2) - \frac{1}{\omega+{\rm i}\tau^{-1}} 2{\bm e}_\| ({\bm k}\cdot{\bm e}_\|) \right),\\ 
&{\bm j}_{\rm DC}^{(2)}({\bm r})=\frac{e^3v_{\rm F}^2}{2\pi\hbar^2}\frac{{\bm k}}{\omega(\omega^2+\tau^{-2})}|{\bm e}_\||^2,\label{Eq_j2DC_freestanding}\\
&{\bm e}_\|=-t_{\rm p}\frac{\gamma}{q}\hat{\bm k}_\| + t_{\rm s}\hat{\bm k}_\perp, 
\end{align}
according to Eqs. (\ref{Eq_SHG_current}) and (\ref{Eq_PDE_current}). 
We should note that these currents vanish for the normal incidence (${\bm k}=0$). The PDE current ${\bm j}_{\rm DC}^{(2)}$ is obviously uniform.

The electric field of the reflected SH wave becomes 
\begin{align}
&{\bm E}_{2\omega}({\bm x})={\rm e}^{2{\rm i}({\bm k}\cdot{\bm r}+\gamma z)} (t_{\rm p}^+{\bm p}^+ + t_s^+{\bm s}),\\
&t_{\rm p}^+=-\frac{c\mu_0}{2+2\frac{\gamma}{q}\alpha_e\tilde{\sigma}_{xx}(2\omega)} 
\hat{\bm k}_\|\cdot{\bm j}_{2\omega,2{\bm k}}^{(2)},\label{Eq_SHG_freestanding_P}\\
&t_{\rm s}^+=-\frac{c\mu_0\frac{q}{\gamma}}{2+2\frac{q}{\gamma}\alpha_e\tilde{\sigma}_{xx}(2\omega)} \hat{\bm k}_\perp\cdot{\bm j}_{2\omega,2{\bm k}}^{(2)}. \label{Eq_SHG_freestanding_S}
\end{align}
The SHG intensity in the specular reflected direction is given by $\sqrt{ |t_{\rm p}^+|^2+|t_{\rm s}^+|^2}$.

\section{Numerical results and discussions}

Let us consider the doped graphene whose Fermi level lies at 
$E_{\rm F}=0.4$ eV from the Dirac point.  The relaxation time is taken as 
$\tau=0.4$ ps estimated from a typical DC mobility of $10^4$ [cm$^2$/(Vs)] \cite{koppens2011graphene}.  The grating consists of the parallel array of identical circular rods arranged in the triangular lattice with lattice constant $a$ of 1 $\mu$m.  The radius, height, and dielectric constant of the rods are taken to be $0.3a$, $0.6a$, and 12, respectively. The grating is supported by semi-infinite high-index substrate of dielectric constant of 12.  The graphene is placed just above the grating with distance $0.01a$. Using the method presented in Ref. \onlinecite{ochiai2015spatially}, the periodic modulation of the linear conductivity and plasmonic band structure are obtained.

\begin{figure}
\centerline{
\includegraphics[width=0.45\textwidth]{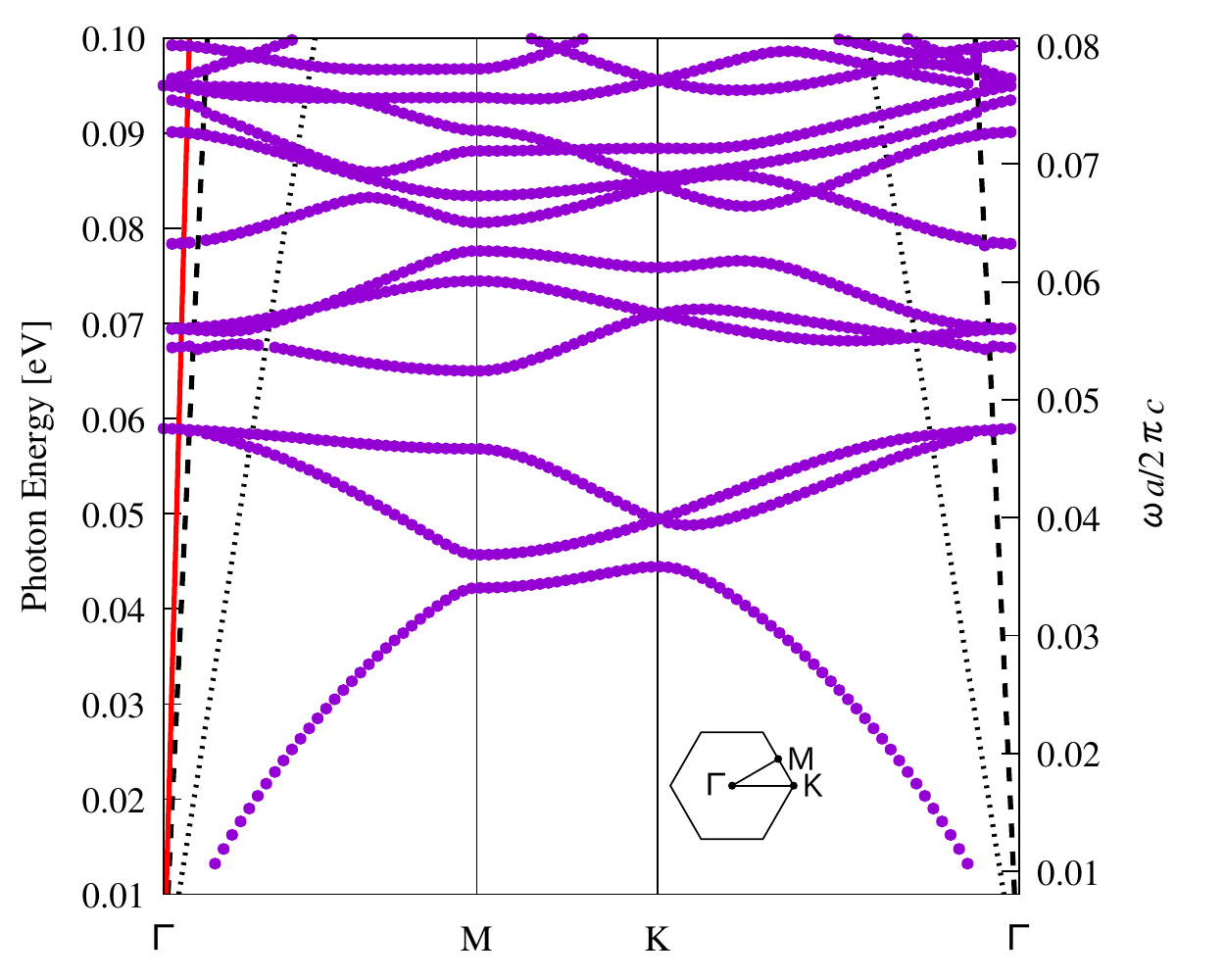}
}
\caption{\label{Fig_band} Plasmonic band structure of the doped graphene with $E_{\rm F}$=0.4 eV placed on a triangular-lattice diffraction grating with lattice constant $a=1$ $\mu$m and thickness $d=0.6a$, composed of circular rods with radius $0.3a$ and dielectric constant $\epsilon_a=12$. The background dielectric constant $\epsilon_b=1$. Those of the upper region (air) and lower region (substrate) are  taken to be $\epsilon_u=1$ and $\epsilon_l=12$, respectively. The distance between the grating and graphene is $0.01a$ ($z_c=0.31a$) [see Fig. \ref{Fig_setup}]. 
Dashed and dotted lines are the light lines of air and substrate, respectively. Solid line (red) stands for the dispersion line of the incident light, $|{\bm k}|=(\omega/c) \sin\theta_{\rm in}$ with $\theta_{\rm in}=\pi/5$, in the $\Gamma$M direction.   
The dissipation is taken to be zero ($\tau=\infty$) for the well-defined band structure.  The inset shows the first Brillouin zone. 
}
\end{figure}

Figure \ref{Fig_band} show the plasmonic band structure of the system. 
The band structure is basically the zone folding of the plasmon polariton of the free-standing graphene. However, the presence of the grating and substrate strongly modulates the plasmon-polariton dispersion, forming the plasmonic band gap structure. 
In particular, the plasmonic band gap is found around $\omega a/2\pi c=0.05$.
This value is much smaller than typical photonic band-gap frequencies of dielectric photonic crystals, which is typically around  $\omega a/2\pi c=0.5$ \cite{Joannopoulos-PC-book}.

\begin{figure}
\centerline{
\includegraphics[width=0.45\textwidth]{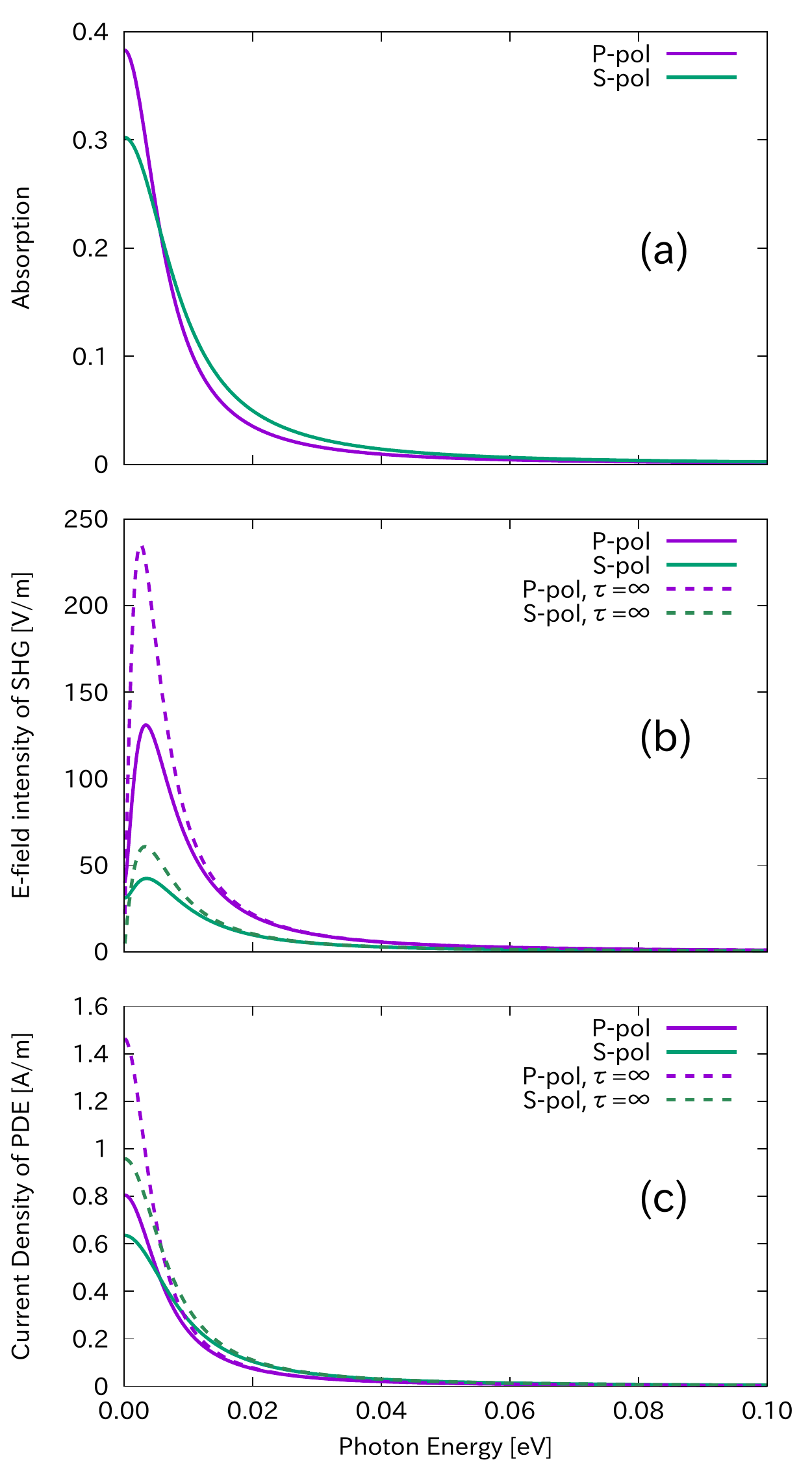}
}
\caption{\label{Fig_freestanding} (a) Optical absorption, (b) electric-field intensity $\sqrt{|t_{p}^+|^2+|t_{s}^+|^2}$ of the SHG light, and (c) direct current density ${\bm j}_{\rm DC}^{(2)}$ in the direction parallel to ${\bm k}$, of the doped graphene with $E_{\rm F}$=0.4 eV, as a function of photon energy of incident light.  The relaxation time of the graphene is taken to be $\tau$=0.4 ps. Incident angle is fixed to $\theta_{\rm in}=\pi/5$ (normal incidence corresponds to $\theta_{\rm in}=0$). The electric-field intensity of the incident light is taken to be $10^6$ [V/m]. Both P- and S-polarized incident lights are considered. In (b) and (c), the results of $\tau=\infty$ are also plotted for comparison. 
}
\end{figure}

Before considering the SHG and PDE in the doped graphene on the grating, we sketch the results of the free-standing graphene for comparison. 
Figure \ref{Fig_freestanding} shows the absorption, SHG and PDE spectra of the free-standing doped graphene under an incident plane-wave light.

The absorption becomes large at small photon energies, where the linear conductivity of Eq. (\ref{Eq_drude}) approaches to be real, and thus the graphene becomes highly dissipative.
 We should note that the well-known optical absorption of 2.3\% of the graphene \cite{nair2008fine} emerges above the interband-transition threshold $2E_{\rm F}$, which is far outside the energy range of Fig.~\ref{Fig_freestanding}.

The SHG spectrum exhibits a peak near the zero energy, but the position of this peak moves by changing the incident angle (not shown). The peak thus merely reflects the absorption peak at zero energy, affected  by energy scale of $\hbar/\tau$, which is the unique energy scale other than $E_{\rm F}$. Note that we are concerned with the energy range much below $2E_{\rm F}$, otherwise the interband transition is relevant and the classical approach of the nonlinear conductivity becomes ineffective. Besides, the PDE spectrum monotonically decreases with increasing frequency. We also note that even in the clean limit ($\tau=\infty$), the SHG and PDE spectra of the free-standing graphene do not change so much.

In this setup, the graphene plasmon is not relevant to these spectra. The incident light cannot excite the graphene plasmon, because its dispersion is outside the light cone and thus frequency and momentum of the incident light can not match with those of the graphene plasmon.   
We also note that the SHG and PDE  vanish at the normal incidence, because the nonlinear currents involves the spatial derivative of the electric field in plane.  
In the normal incidence, the electric field is uniform in plane, and thus the nonlinear current vanishes.

\begin{figure}
\centerline{
\includegraphics[width=0.45\textwidth]{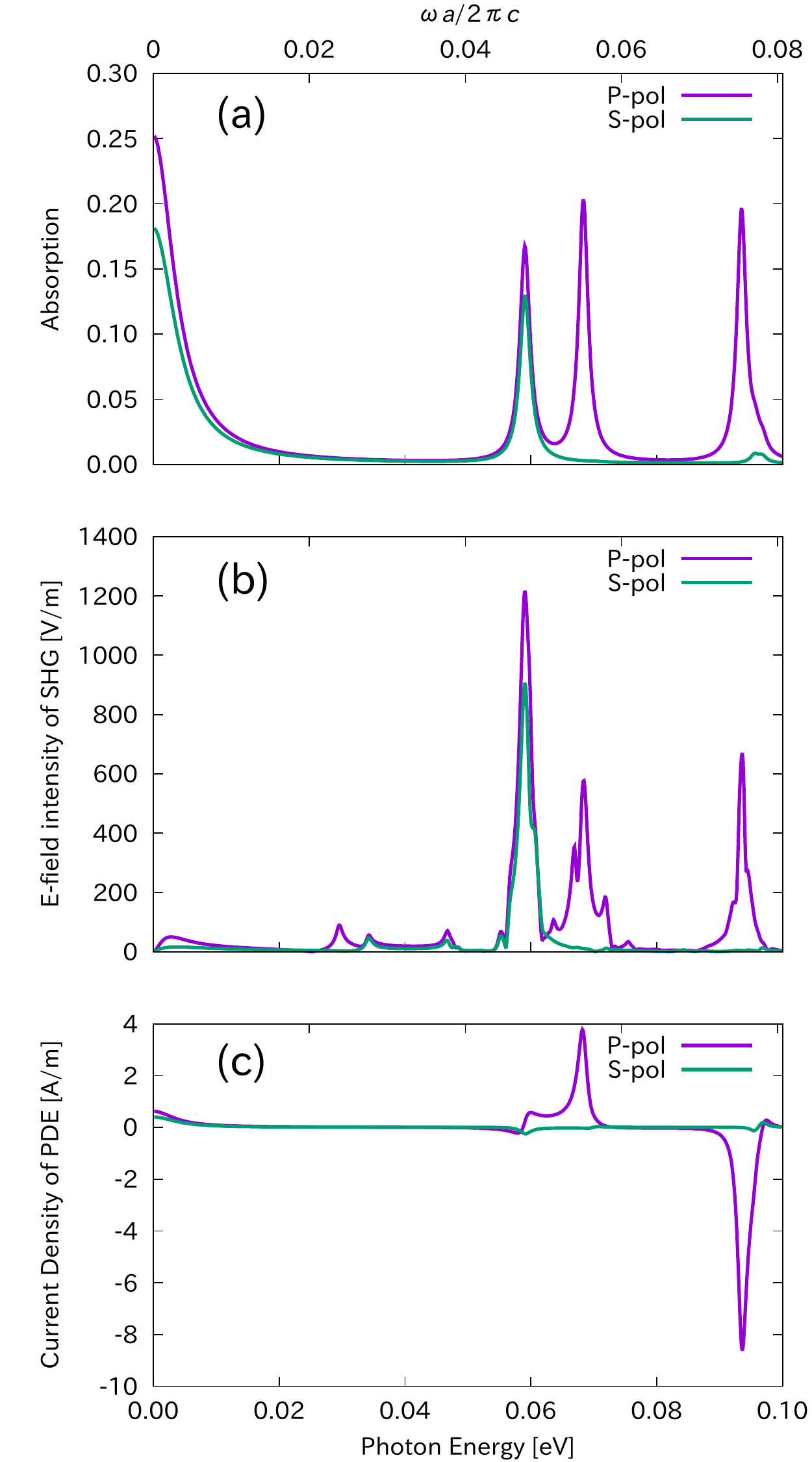}
}
\caption{\label{Fig_graphene_grating}(a) Optical absorption, (b) electric-field intensity $\sqrt{|t_{p0}^+|^2+|t_{s0}^+|^2}$, and (c) spatially averaged direct current ${\bm j}_{{\rm DC},0}^{(2)}$ in the $\Gamma$M direction  of the graphene-grating system, as a function of the photon energy of the incident light.  The incident light is coming from the top of the structure with incident angle $\theta_{\rm in}=\pi/5$ in the $\Gamma$M direction. Relaxation time $\tau$ is taken to be 0.4 ps. Two polarizations of the incident light are examined.  }
\end{figure}

If the graphene is put on the grating, the results change drastically.  
Figure \ref{Fig_graphene_grating} show the absorption, SHG, and PDE spectra in the graphene-grating system depicted in Fig. \ref{Fig_setup}. 
Here, the incident angle is tilted to the $\Gamma$M direction of the triangular-lattice grating.

The absorption spectrum has three marked peaks at $\hbar\omega=$0.059, 0.063, and  0.094 eV, other than the zero-energy peak found also in Fig.~\ref{Fig_freestanding} (a).  These peaks correspond to the intersection points between the dispersion line ${\bm k}=(\omega/c)\sin\theta_{\rm in}(\cos\phi_{\rm in},\sin\phi_{\rm in})$ of the incident light and the plasmonic band-stricture curves in Fig.~\ref{Fig_band}. 
We should note the intersection points around $\hbar\omega=$0.08 and 0.09 eV in Fig. \ref{Fig_band} do not emerge in the absorption spectra. The relevant bands are  non-degenerate at $\Gamma$, so that they are the so-called uncoupled modes there due to a symmetry mismatch \cite{PhysRevB.58.6920,Ochiai:S::63:p125107:2001}.  Reflecting this uncoupled nature, the coupling between the incident light and the plasmon mode is very small at nonzero ${\bm k}$. 
We also note the photonic band modes are selectively excited by the polarization of the incident light, because the plasmon mode in the $\Gamma$M axis 
are classified according to the parity with respect to the axis. The P- and S-polarized incident light has the even and odd parities, respectively, so that the P-polarized incident light, for instance,  cannot excite the odd-parity modes.

The SHG spectrum shown in Fig. \ref{Fig_graphene_grating} (b) exhibits many peaks. We emphasize here that the difference in the vertical-axis scale between 
Fig. \ref{Fig_graphene_grating} (b) and Fig. \ref{Fig_freestanding} (b).  
At marked peaks of $\hbar\omega=0.059$ and 0.094 eV, the enhancement factors relative to the free-standing graphene are about 470 and 640, respectively. These factors are independent of the intensity of the incident light, but increase with increasing relaxation time $\tau$.

Such an enhancement comes from the phase matching to the graphene plasmon polariton. 
Many peaks in the SHG spectrum are due to a two-fold enhancement of the FH and SH waves. Namely,  
the enhancement of the SHG occurs when the phase matching of either the FH and/or SH wave are met. The incident FH wave yields a large electric-field intensity  in the graphene layer, when incident light of $(\omega,{\bm k})$ is phase-matched with the plasmon dispersion. This matching  results in large nonlinear current of ${\bm j}_{2\omega}^{(2)}$. This is the first enhancement. The second enhancement comes from the current-induced radiation of the SH wave. Even if the nonlinear current itself is not enhanced, the radiation is enhanced when the SH wave of $(2\omega,2{\bm k})$ is phase-matched. By tuning the plasmonic band structure, it is possible to obtain the double enhancement of both the FH and SH waves.
 A similar two-fold enhancement is well known in the surface-enhanced Raman scattering \cite{PhysRevE.62.4318}.

The PDE spectrum of Fig. \ref{Fig_graphene_grating} (c) again shows the excitation of the graphene plasmon polariton of the FH wave.  
Compared to the free-standing graphene, the current density is strongly enhanced. At the main peak of $\hbar\omega=0.094$ eV, the enhancement factor relative to the free-standing graphene is about 2300. Again, this factor is independent of the intensity of the incident light, but increase with increasing $\tau$. 
By the mirror symmetry of the grating structure,  
the average direct current flows along the $\Gamma$M axis. However, its orientation can be both forward and backward, depending on the plasmon-mode configuration under consideration.
This is not the case for the free-standing graphene. There, the current density is always directed to ${\bm k}$ of the incident light with the sign-definite coefficient [see Eq. (\ref{Eq_j2DC_freestanding})].

\begin{figure}
\centerline{
\includegraphics[width=0.45\textwidth]{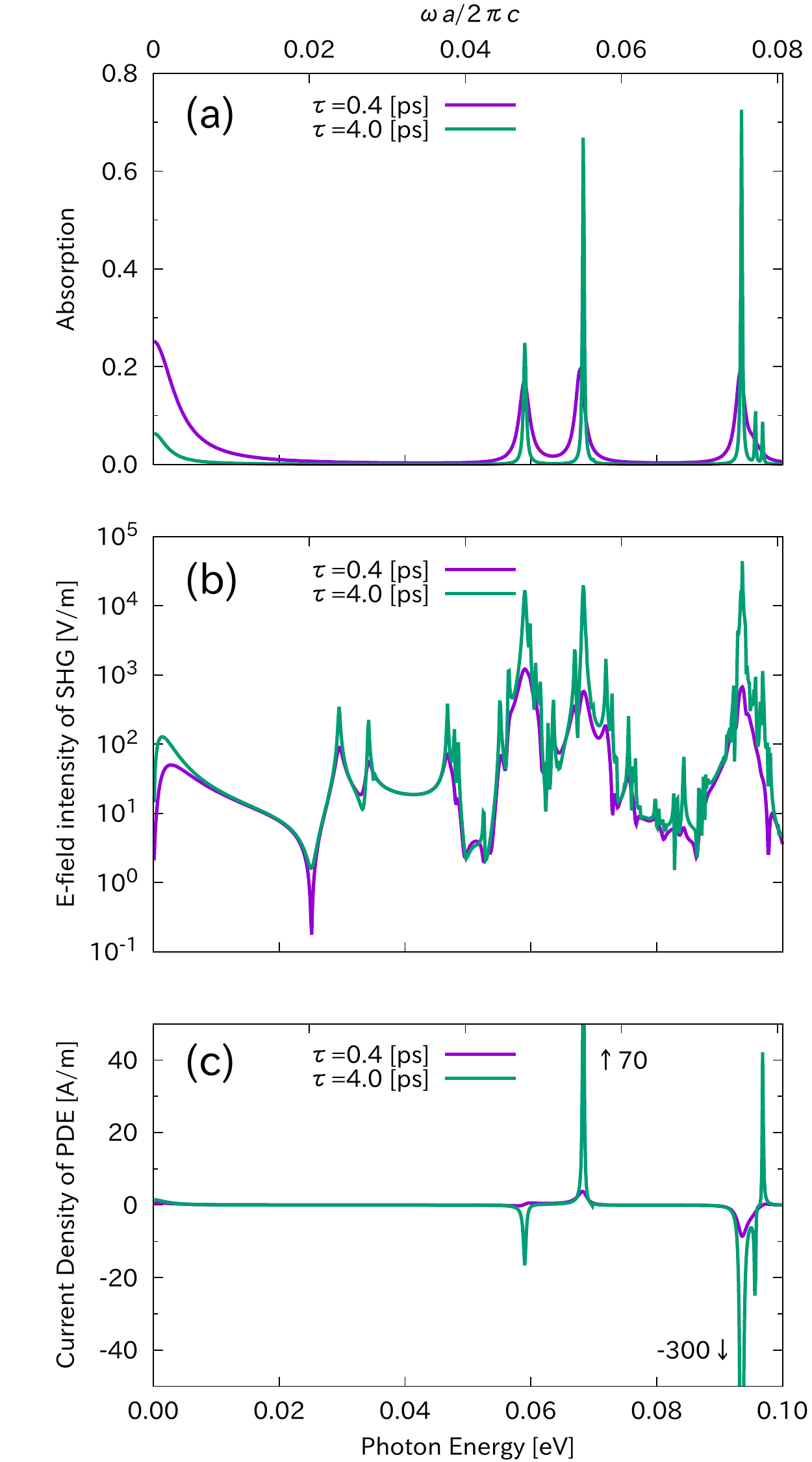}
}
\caption{\label{Fig_taudepend} Relaxation time dependence in (a) optical absorption, (b) electric-field intensity $\sqrt{|t_{p0}^+|^2+|t_{s0}^+|^2}$, and (c) spatially averaged direct current ${\bm j}_{{\rm DC},0}^{(2)}$ in the $\Gamma$M direction  of the graphene-grating system.  The P-polarized incident light is coming from the top of the structure with incident angle $\theta_{\rm in}=\pi/5$ in the $\Gamma$M direction.  
}
\end{figure}

To see how the relaxation time affects the SHG and PDE spectra, we show in Fig. \ref{Fig_taudepend}, the spectra for ten-times longer relaxation time (4 ps) of the graphene.  
Such a value of $\tau$ is available for cleaner sample with high mobility \cite{bolotin2008ultrahigh}.   
The spectra become sharper around peak frequencies, while the background spectra do not change so much. There is a striking contrast between unmodulated and modulated graphene, regarding this feature. In the unmodulated graphene, even at $\tau=\infty$ (no dissipation), the SHG and PDE spectra do not change so much from those at modest $\tau$, as shown in Fig. \ref{Fig_freestanding}. The enhancement factor of the SHG and PDE is roughly one-order magnitude higher than in the spectra of the modest $\tau$ of 0.4 ps.

\begin{figure}
\begin{center}
\includegraphics[width=0.45\textwidth]{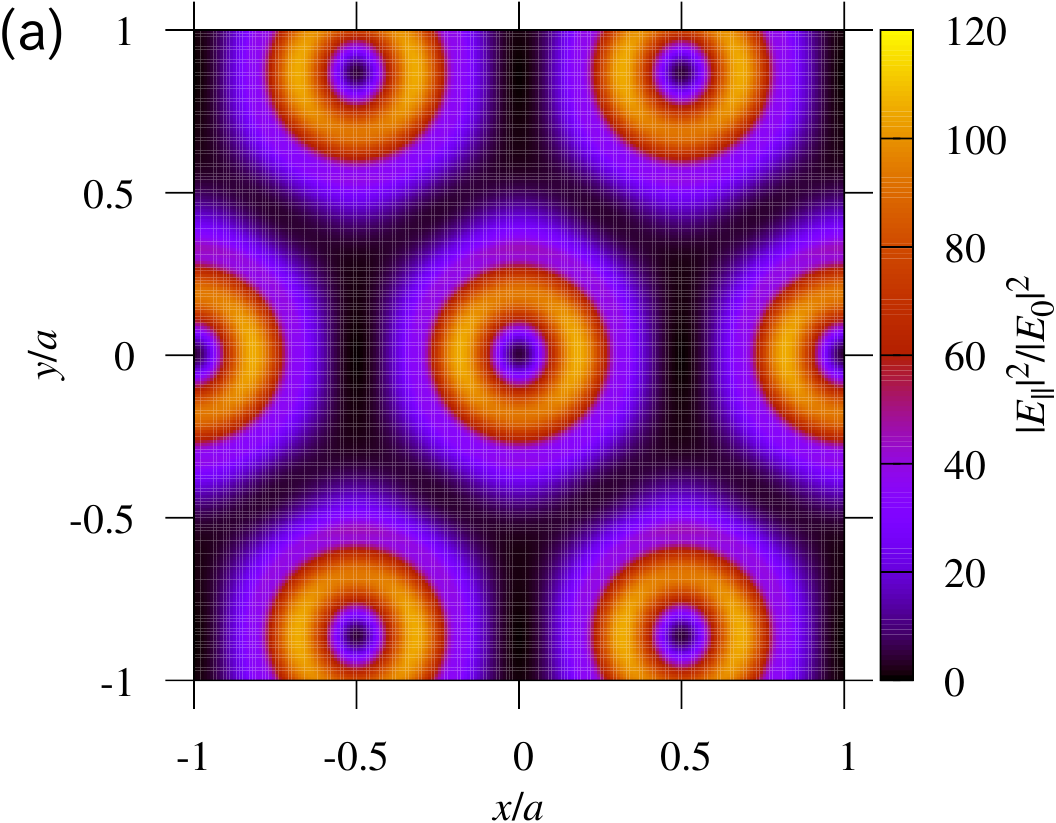}
\includegraphics[width=0.45\textwidth]{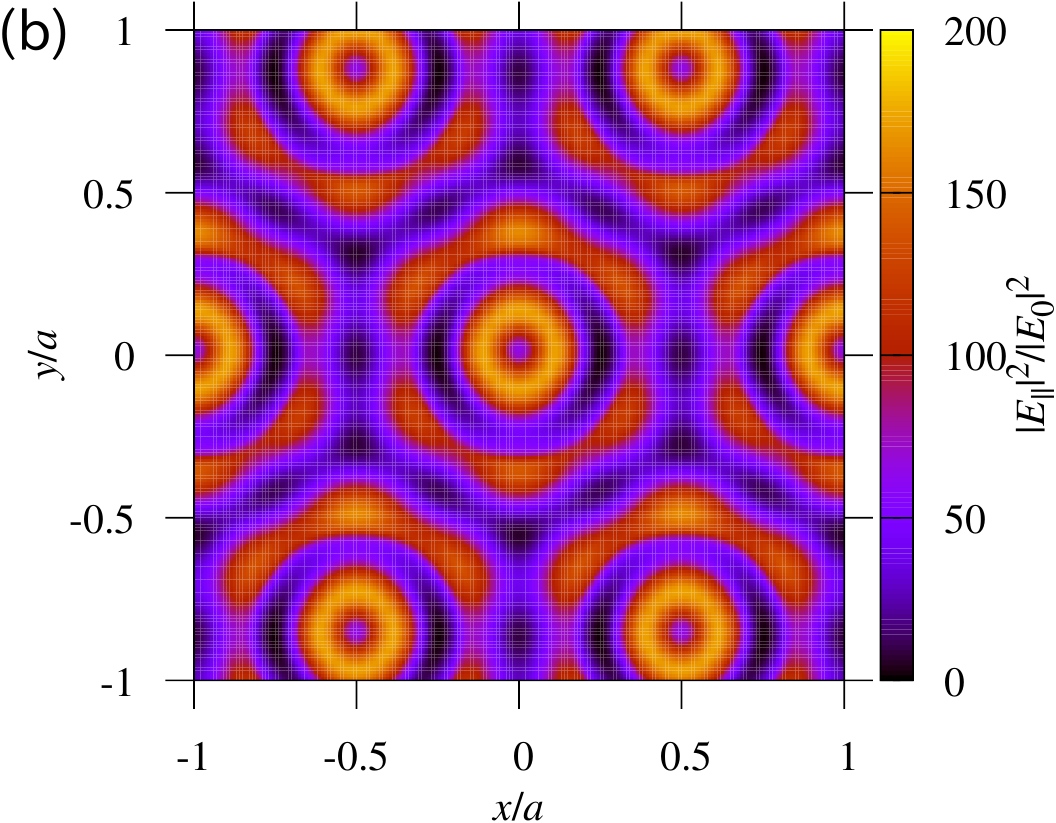}
\end{center}
\caption{\label{Fig_Econf_FH} Electric-field profile (of the FH wave) in the doped graphene on the diffraction grating. The incident light is P-polarized. Photon energy is taken to be (a) 0.063 and (b) 0.094 eV, which correspond to the second and third marked peaks in Fig. \ref{Fig_graphene_grating} (a). 
}
\end{figure}

The field configuration of the graphene plasmon at $\hbar\omega=$0.063 and 0.094 eV, which is responsible to the enhanced SHG and PDE,  is shown in Fig. \ref{Fig_Econf_FH}.
The electric field intensity in the graphene layer is strongly modulated with a triangular-lattice pattern by the grating. In the unmodulated graphene, the electric-field intensity is uniform.  
This intensity pattern, however, is not sufficient to understand the PDE.

\begin{figure}
\begin{center}
\includegraphics[width=0.45\textwidth]{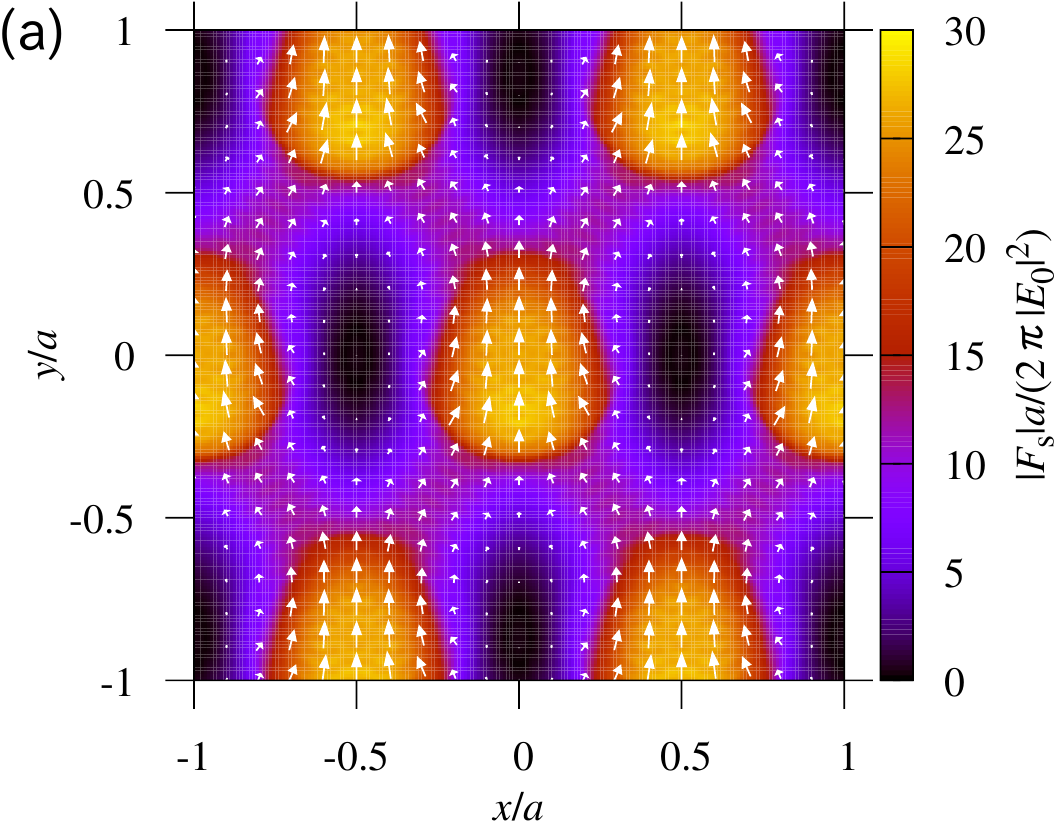}
\includegraphics[width=0.45\textwidth]{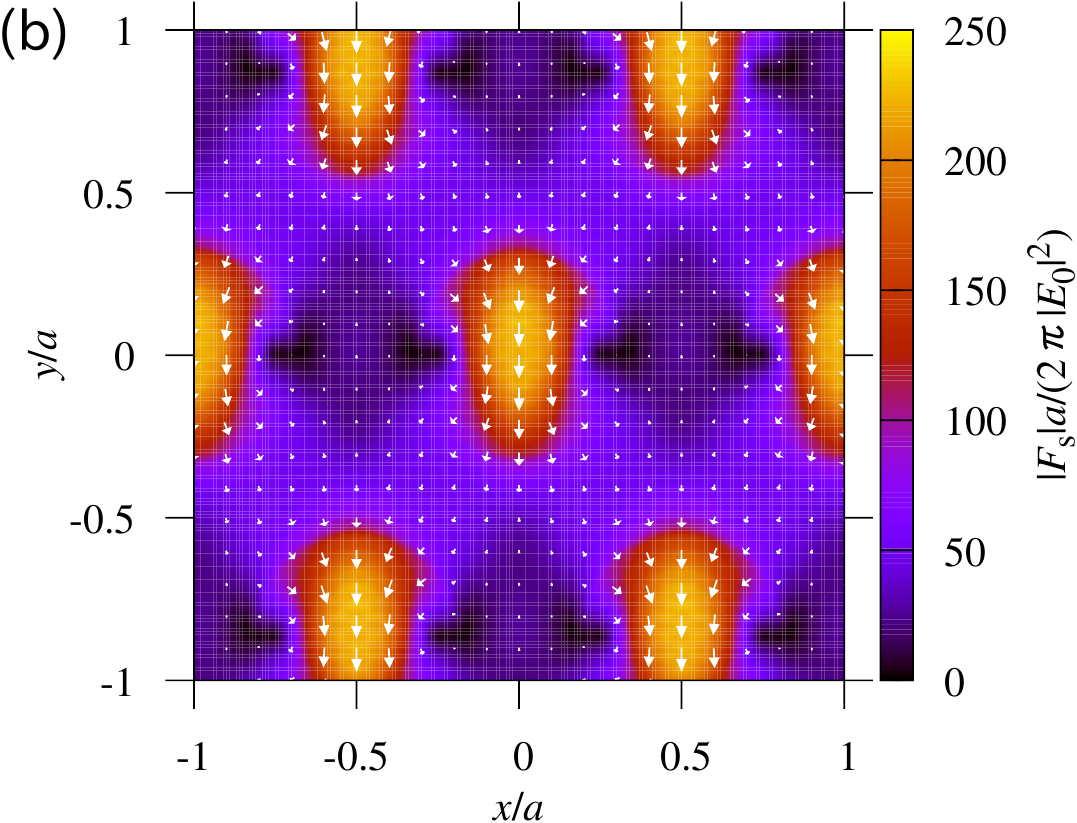}
\end{center}
\caption{\label{Fig_force_FH} The scattering-force term $({\bm F}_s)_i=\Im[({\bm E}_\omega)_k^*\partial_i({\bm E}_\omega)_k]$ in the doped graphene on the diffraction grating. The incident light is P-polarized. Photon energy is taken to be (a) 0.063 and (b) 0.094 eV, which correspond to the second and third marked peaks in Fig. \ref{Fig_graphene_grating} (a). 
}
\end{figure}

As we saw in Figs. \ref{Fig_graphene_grating} (c) and \ref{Fig_taudepend} (c), the PDE current can flow both forward and backward directions to the incident light, depending on the plasmon mode concerned.   
The difference in sign of the averaged PDE current is not related to the electric field intensity $|{\bm E}_\omega|^2$ or its spatial derivative (relevant to the the gradient force), but is related to the scattering force (for a polarizable probe particle). 
This property can be understood by reconsidering Eq. (\ref{Eq_PDE_current}). By dropping the total derivative term, which does not contribute to the average current, the direct current density becomes  
\begin{align}
[{\bm j}_{\rm DC}^{(2)}({\bm r})]_i\to \frac{e^3 v_{\rm F}^2}{2\pi\hbar^2}\frac{1}{\omega(\omega^2 + \tau^{-2} )} \Im[({\bm E}_\omega)_k^*\partial_i ({\bm E}_\omega)_k].  
\end{align}
The factor $F_i^{({\rm s})}=\Im[({\bm E}_\omega)_k^*\partial_i ({\bm E}_\omega)_k]$ corresponds to the scattering force \cite{chaumet2000tat,PhysRevLett.102.113602,onoda2009designing}.  
This factor is plotted in Fig. \ref{Fig_force_FH}. 
Obviously, the force flows in the opposite directions between the two modes, resulting in the opposite sign of the average direct current. In this way, characters of the PDE are strongly tied to the scattering force.

Finally, we should comment on effects of finite temperature and thickness in the doped graphene. In the present paper, we assume zero temperature and infinitely thin conducting sheet for the graphene. 
In a real specimen, however, temperature and thickness are finite in the graphene sheet. These effects are fairly neglected as follows. 

The former property can be implemented by using the Fermi-distribution function at finite temperature $T$, instead of the step-like distribution at zero temperature, in the Kubo formula of the optical conductivity. In the parameter range of interest, this effect merely causes a small shift in the conductivity as $\sigma_\omega^{(1)}|_{T\ne 0}\simeq \sigma_\omega^{(1)}|_{T=0}[1+2k_{\rm B}T/E_{\rm F}\exp(-E_{\rm F}/k_{\rm B}T)]$ \cite{hanson2008dyadic}. At $E_{\rm F}=$0.4 eV and room temperature, the shift is of order $10^{-8}$. 

We can also implement effects of finite thickness by using a thin conducting film with thickness $d_{\rm gra}$. The permittivity of the film should be uniaxial with $\tensor{\epsilon}={\rm diag}(\epsilon_\|,\epsilon_\|,\epsilon_\perp)$. The in-plane permittivity is $\epsilon_\|=1+{\rm i}\sigma_\omega^{(1)}/(\epsilon_0\omega d_{\rm gra})$. The out-of-plane permittivity is  $\epsilon_\perp=1$.  Using this model, we can evaluate various spectra including the absorption and SHG.   The results with $d_{\rm gra}=$ 0.3 to 0.5 nm, which are typical thickness values of the graphene, are almost the same as those with the infinitely thin conducting sheet. This is because the length scale of $d_{\rm gra}$ is much smaller than relevant length scales, particularly the spatial decay length $1/\sqrt{{\bm k}^2-(\omega/c)^2}$ of the graphene plasmon polariton. 
\color{black}

\section{Conclusion}

We have shown theoretically the enhancement of the SHG and PDE in a doped graphene placed on a two-dimensional diffraction grating. The relevant second-order optical conductivity is obtained in a semi-classical approach with a corresponding principle to the Dirac spectrum of the graphene. The SHG and PDE are described with a generalization of the rigorous coupled-wave analysis formalism taking account of infinitely thin electrically conducting layer of the doped graphene. 
 The results of the SHG and PDE spectra are compared with those in the free-standing graphene. By exciting the graphene plasmon modulated by the grating, the enhancement factor of the SHG and PDE reaches to 640 and 2300, relative to the free-standing doped graphene, for the system with a conservatively estimated value of the relaxation time.   Longer relaxation time yields higher enhancement factor on the tradeoff of narrower resonance peak of the graphene plasmon. The PDE current can flow in both the forward and backward directions to the incident light, and is related to the scattering force for the plasmon mode concerned.  
This high enhancement of the SHG and PDE together with the mode-sensitive direction of the PDE current opens up a new application of selective frequency conversion and THz photo-detection via the modulated graphene.

\section*{acknowledgments}
This work was supported by JSPS KAKENHI Grant No. 26390013.  


\begin{thebibliography}{10}
\newcommand{\enquote}[1]{``#1''}

\bibitem{RevModPhys.81.109}
A.~H. Castro~Neto, F.~Guinea, N.~M.~R. Peres, K.~S. Novoselov, and A.~K. Geim,
  \enquote{The electronic properties of graphene,} Rev. Mod. Phys. \textbf{81},
  109--162 (2009).

\bibitem{novoselov2005tdg}
K.~S. Novoselov, A.~K. Geim, S.~V. Morozov, D.~Jiang, M.~I. Katsnelson, I.~V.
  Grigorieva, S.~V. Dubonos, and A.~A. Firsov, \enquote{{Two-dimensional gas of
  massless Dirac fermions in graphene},} Nature \textbf{438}, 197--200 (2005).

\bibitem{zhang2005experimental}
Y.~Zhang, Y.-W. Tan, H.~L. Stormer, and P.~Kim, \enquote{Experimental
  observation of the quantum hall effect and berry's phase in graphene,} Nature
  \textbf{438}, 201--204 (2005).

\bibitem{PhysRevB.82.201402}
K.~L. Ishikawa, \enquote{Nonlinear optical response of graphene in time
  domain,} Phys. Rev. B \textbf{82}, 201,402 (2010).

\bibitem{hendry2010coherent}
E.~Hendry, P.~J. Hale, J.~Moger, A.~K. Savchenko, and S.~A. Mikhailov,
  \enquote{Coherent nonlinear optical response of graphene,} Phys. Rev. Lett.
  \textbf{105}, 097,401 (2010).

\bibitem{yazyev2010emergence}
O.~V. Yazyev, \enquote{Emergence of magnetism in graphene materials and
  nanostructures,} Rep. Prog. Phys. \textbf{73}, 056,501 (2010).

\bibitem{han2014graphene}
W.~Han, R.~K. Kawakami, M.~Gmitra, and J.~Fabian, \enquote{Graphene
  spintronics,} Nat. Nanotech. \textbf{9}, 794--807 (2014).

\bibitem{bludov2013primer}
Y.~V. Bludov, A.~Ferreira, N.~M.~R. Peres, and M.~I. Vasilevskiy, \enquote{A
  primer on surface plasmon-polaritons in graphene,} Int. J. Mod. Phys. B
  \textbf{27}, 1341,001 (2013).

\bibitem{koppens2011graphene}
F.~H.~L. Koppens, D.~E. Chang, and F.~J. Garc\'ia~de Abajo, \enquote{Graphene
  plasmonics: a platform for strong light--matter interactions,} Nano Lett.
  \textbf{11}, 3370--3377 (2011).

\bibitem{fahrenfort1961attenuated}
J.~Fahrenfort, \enquote{Attenuated total reflection: A new principle for the
  production of useful infra-red reflection spectra of organic compounds,}
  Spectrochim. Acta \textbf{17}, 698--709 (1961).

\bibitem{dakss1970grating}
M.~L. Dakss, L.~Kuhn, P.~F. Heidrich, and B.~A. Scott, \enquote{Grating coupler
  for efficient excitation of optical guided waves in thin films,} Appl. Phys.
  Lett. \textbf{16}, 523--525 (1970).

\bibitem{ochiai2015spatially}
T.~Ochiai, \enquote{Spatially periodic modulation of optical conductivity in
  doped graphene by two-dimensional diffraction grating,} J. Opt. Soc. Am. B
  \textbf{32}, 701--707 (2015).

\bibitem{manzoni2015second}
M.~T. Manzoni, I.~Silveiro, F.~J. Garc\'{\i}a~de Abajo, and D.~E. Chang,
  \enquote{Second-order quantum nonlinear optical processes in single graphene
  nanostructures and arrays,} New J. Phys. \textbf{17}, 083,031 (2015).

\bibitem{PhysRevB.92.161406}
D.~A. Smirnova, A.~E. Miroshnichenko, Y.~S. Kivshar, and A.~B. Khanikaev,
  \enquote{Tunable nonlinear graphene metasurfaces,} Phys. Rev. B \textbf{92},
  161,406 (2015).

\bibitem{PhysRevB.81.165441}
M.~V. Entin, L.~I. Magarill, and D.~L. Shepelyansky, \enquote{Theory of
  resonant photon drag in monolayer graphene,} Phys. Rev. B \textbf{81},
  165,441 (2010).

\bibitem{ivchenko2012photoinduced}
E.~Ivchenko, \enquote{Photoinduced currents in graphene and carbon nanotubes,}
  Phys. Status solidi (b) \textbf{249}, 2538--2543 (2012).

\bibitem{PhysRevB.90.241416}
P.~A. Obraztsov, N.~Kanda, K.~Konishi, M.~Kuwata-Gonokami, S.~V. Garnov, A.~N.
  Obraztsov, and Y.~P. Svirko, \enquote{Photon-drag-induced terahertz emission
  from graphene,} Phys. Rev. B \textbf{90}, 241,416 (2014).

\bibitem{glazov2014high}
M.~M. Glazov and S.~D. Ganichev, \enquote{High frequency electric field induced
  nonlinear effects in graphene,} Phys. Rep. \textbf{535}, 101--138 (2014).

\bibitem{PhysRevB.92.235307}
J.~L. Cheng, N.~Vermeulen, and J.~E. Sipe, \enquote{Numerical study of the
  optical nonlinearity of doped and gapped graphene: From weak to strong field
  excitation,} Phys. Rev. B \textbf{92}, 235,307 (2015).

\bibitem{PhysRevB.24.849}
A.~Wokaun, J.~G. Bergman, J.~P. Heritage, A.~M. Glass, P.~F. Liao, and D.~H.
  Olson, \enquote{Surface second-harmonic generation from metal island films
  and microlithographic structures,} Phys. Rev. B \textbf{24}, 849--856 (1981).

\bibitem{nahata2003enhanced}
A.~Nahata, R.~A. Linke, T.~Ishi, and K.~Ohashi, \enquote{Enhanced nonlinear
  optical conversion from a periodically nanostructured metal film,} Opt. Lett.
  \textbf{28}, 423--425 (2003).

\bibitem{airola2005second}
M.~Airola, Y.~Liu, and S.~Blair, \enquote{Second-harmonic generation from an
  array of sub-wavelength metal apertures,} J. Opt. A: Pure Appl. Opt.
  \textbf{7}, S118 (2005).

\bibitem{PhysRevLett.103.103906}
T.~Hatano, T.~Ishihara, S.~G. Tikhodeev, and N.~A. Gippius, \enquote{Transverse
  photovoltage induced by circularly polarized light,} Phys. Rev. Lett.
  \textbf{103}, 103,906 (2009).

\bibitem{kurosawa2012surface}
H.~Kurosawa and T.~Ishihara, \enquote{Surface plasmon drag effect in a
  dielectrically modulated metallic thin film,} Opt. express \textbf{20},
  1561--1574 (2012).

\bibitem{noginova2013plasmon}
N.~Noginova, V.~Rono, F.~J. Bezares, and J.~D. Caldwell, \enquote{Plasmon drag
  effect in metal nanostructures,} New J. Phys. \textbf{15}, 113,061 (2013).

\bibitem{PhysRevLett.117.083901}
H.~Kurosawa, K.~Sawada, and S.~Ohno, \enquote{Photon drag effect due to berry
  curvature,} Phys. Rev. Lett. \textbf{117}, 083,901 (2016).

\bibitem{kragler1980dielectric}
R.~Kragler and H.~Thomas, \enquote{Dielectric function in the relaxation-time
  approximation generalized to electronic multiple-band systems,} Z. Phys. B
  Cond. Mat. \textbf{39}, 99--107 (1980).

\bibitem{Noponen1994}
E.~Noponen and J.~Turunen, \enquote{Eigenmode method for electromagnetic
  synthesis of diffractive elements with three-dimensional profiles,} J. Opt.
  Soc. Am. A \textbf{11}, 2494--2502 (1994).

\bibitem{Li1997}
L.~Li, \enquote{New formulation of the fourier modal method for crossed
  surface-relief gratings,} J. Opt. Soc. Am. A \textbf{14}, 2758--2767 (1997).

\bibitem{Tikhodeev:Y:M:G:I::66:p045102:2002}
S.~G. Tikhodeev, A.~L. Yablonskii, E.~A. Muljarov, N.~A. Gippius, and
  T.~Ishihara, \enquote{Quasiguided modes and optical properties of photonic
  crystal slabs,} Phys. Rev. B \textbf{66}, 045,102 (2002).

\bibitem{Joannopoulos-PC-book}
J.~D. Joannopoulos, R.~D. Meade, and J.~N. Winn, \emph{Photonic Crystals}
  (Princeton University Press, Princeton, 1995).

\bibitem{nair2008fine}
R.~R. Nair, P.~Blake, A.~N. Grigorenko, K.~S. Novoselov, T.~J. Booth,
  T.~Stauber, N.~M. Peres, and A.~K. Geim, \enquote{Fine structure constant
  defines visual transparency of graphene,} Science \textbf{320}, 1308--1308
  (2008).

\bibitem{PhysRevB.58.6920}
H.~Miyazaki and K.~Ohtaka, \enquote{Near-field images of a monolayer of
  periodically arrayed dielectric spheres,} Phys. Rev. B \textbf{58},
  6920--6937 (1998).

\bibitem{Ochiai:S::63:p125107:2001}
T.~Ochiai and K.~Sakoda, \enquote{Dispersion relation and optical transmittance
  of a hexagonal photonic crystal slab,} Phys. Rev. B \textbf{63}, 125,107
  (2001).

\bibitem{PhysRevE.62.4318}
H.~Xu, J.~Aizpurua, M.~K\"all, and P.~Apell, \enquote{Electromagnetic
  contributions to single-molecule sensitivity in surface-enhanced raman
  scattering,} Phys. Rev. E \textbf{62}, 4318--4324 (2000).

\bibitem{bolotin2008ultrahigh}
K.~I. Bolotin, K.~J. Sikes, Z.~Jiang, M.~Klima, G.~Fudenberg, J.~Hone, P.~Kim,
  and H.~L. Stormer, \enquote{Ultrahigh electron mobility in suspended
  graphene,} Solid State Commun. \textbf{146}, 351--355 (2008).

\bibitem{chaumet2000tat}
P.~C. Chaumet and M.~Nieto-Vesperinas, \enquote{{Time-averaged total force on a
  dipolar sphere in an electromagnetic field},} Opt. Lett. \textbf{25},
  1065--1067 (2000).

\bibitem{PhysRevLett.102.113602}
S.~Albaladejo, M.~I. Marqu\'es, M.~Laroche, and J.~J. S\'aenz,
  \enquote{Scattering forces from the curl of the spin angular momentum of a
  light field,} Phys. Rev. Lett. \textbf{102}, 113,602 (2009).

\bibitem{onoda2009designing}
M.~Onoda and T.~Ochiai, \enquote{{Designing Spinning Bloch States in 2D
  Photonic Crystals for Stirring Nanoparticles},} Phys. Rev. Lett.
  \textbf{103}, 033,903 (2009).

\bibitem{hanson2008dyadic}
G.~W. Hanson, \enquote{Dyadic green’s functions and guided surface waves for
  a surface conductivity model of graphene,} J. Appl. Phys. \textbf{103},
  064,302 (2008).

\end{thebibliography}

\end{document}